\documentclass[11pt,acm]{article}
\usepackage[utf8]{inputenc}
\usepackage[english]{babel}
\usepackage{subcaption}
\usepackage{sidecap} 
\usepackage{graphicx}
\usepackage{amsmath}
\usepackage{amssymb}
\usepackage[unicode]{hyperref}
\usepackage[tableposition=top]{caption}
\usepackage{multirow}
\usepackage{tabularx}
\usepackage{textcomp}
\usepackage{gensymb}
\usepackage{graphics}
\usepackage{color}
\usepackage{cleveref}
\usepackage[a4paper, total={6.5in, 8.5in}]{geometry}
\numberwithin{figure}{section}
\numberwithin{table}{section}
\usepackage{titlesec}
\titleformat{\subsection}{\normalfont\itshape}{\thesubsection}{1em}{}
\usepackage{chngcntr}
\counterwithout{figure}{section}
\counterwithout{table}{section}
\usepackage{tcolorbox}

\title{Properties of DBD Plasma Jets using Powered Electrode With and Without Contact with the Plasma}

\author{Fellype do Nascimento, Konstantin Kostov\\ \small{\textit{Faculty of Engineering, UNESP, Guaratingueta, SP, Brazil}} \\ Munemasa Machida \\ \small{\textit{Gleb Wataghin Physics Institute, UNICAMP, Campinas, SP, Brazil}} \\ Alexander Flacker \\ \small{\textit{Renato Archer Center for Information Technology (CTI), Campinas, SP, Brazil}}}

\date{\today}

\begin{document}

\maketitle

\begin{tcolorbox}
\textbf{Note:\\} 
Dear reader,\\If you find this preprint useful and want to cite it, please, refer to the version published in the \textit{IEEE Transactions on Plasma Science} journal:\\F. Nascimento, K. Kostov, M. Machida, A. Flacker, \textit{IEEE Trans. Plasma. Sci.}, Vol. 49, April 2021, pp. 1293-1301 $-$ \href{https://dx.doi.org/10.1109/TPS.2021.3067159}{DOI: 10.1109/TPS.2021.3067159}

\end{tcolorbox}

\begin{abstract}
An experimental investigation comparing the properties of plasma jets in dielectric barrier discharge (DBD) configurations using a powered electrode with and without a dielectric barrier, while keeping a second dielectric barrier over the grounded electrode, is reported in this work. For this purpose, two different power sources were used to produce the plasma jets, with one of them producing a pulsed high-voltage (HV) output and the other one producing a pulse-like HV output, which consists of a damped sine HV waveform. Measurements of plasma parameters were performed for both configurations using argon and helium as working gases. As a result, if the pulsed power source is used, significant differences were found in discharge power ($P_{plasma}$), rotational and vibrational temperatures ($T_r$ and $T_v$, respectively) when switching from one configuration to the other. On the other hand, using the pulse-like HV only the $P_{plasma}$ parameter presented significant differences when switching the electrode's configuration. For the pulsed source it has been observed that despite the remarkable increase in $P_{plasma}$ when changing from the double barrier configuration to the single barrier one, the values obtained for $T_r$ and $T_v$ also increased, but not in the same proportion as the increase in $P_{plasma}$, which suggests a non-linear dependency between temperatures and discharge power in the plasma jet. As an example for application of plasmas in both configurations, tests in an attempt to remove copper films deposited on alumina substrates were performed and, as a result, there was significant material removal only when the powered electrode was in contact with the plasma. As a general conclusion, if higher power is really required for some application it is better to use this configuration.
\end{abstract}


\section{Introduction}
Plasma plumes produced under atmospheric pressure and in open environment generally referred to as Atmospheric Pressure Plasma Jets (APPJs) have been extensively studied in recent years, and a large number of applications for this devices have been developed due to their versatility, easy operation and low cost of implementation compared to low pressure plasmas ~\cite{kogelschatz_dielectric-barrier_2003,lu_guided_2014,brandenburg_dielectric_2017,lu_guided_2018}. The APPJs applications can be in industry, biology and medicine, with the last field receiving great attention, especially due to successful treatments of cancerous tissues and, more recently, for combating viruses, including the Sars-Cov-2 ~\cite{adamovich_2017_2017,reuter_kinpenreview_2018,khlyustova_important_2019,brany_cold_2020, filipic_cold_2020,laroussi_cold_2020}.
Dielectric Barrier Discharge (DBD) is a kind of electrodes arrangement commonly employed to produce APPJs. Its main characteristic is the presence of at least one insulating layer between metallic plates ~\cite{kogelschatz_dielectric-barrier_2003,brandenburg_dielectric_2017}. Among the various DBD configurations, the cylindrical geometry is the most commonly used because it naturally takes advantage of using the gas flow to produce a plasma jet ~\cite{lu_guided_2014}.
Even though already many APPJs studies have been carried out with different arrangements, yet there is no a specific configuration that can be considered ideal, even for specific applications ~\cite{brandenburg_dielectric_2017,lu_guided_2018}. Therefore, research goes on and different geometry configurations and device variations have been explored. It has been a considerable challenge to control plasma jets properties as well as to obtain knowledge about the relationship among different plasma parameters.

Among different ways to arrange the electrodes in cylindrical APPJ reactors, those setups that present powered electrodes at the center of the reactor have been commonly employed. They can have such electrode covered with a dielectric barrier, or not, that is, with the powered electrode in contact with the working gas. Each configuration will produce a plasma jet with different temperatures, power and/or density of reactive species that are created when the plasma interacts with the surrounding ambient air or with a surface ~\cite{adamovich_2017_2017,lu_guided_2018}. The first mentioned configuration (encapsulated powered electrode) presents higher electrical safety when compared to the other (an powered electrode without a dielectric barrier) and, for this reason, it is preferred for medical applications of plasma jets, especially when the treatments are \textit{in vivo}.

Studies regarding the interaction between APPJs and different substrates have shown that the target conductivity is a key parameter that influences not only the treatment results, but can also modify the characteristics of the plasma jet itself ~\cite{prysiazhnyi_properties_2017, simoncelli_experimental_2019,shershunova_features_2019,teschner_investigation_2019}. Works concerning differences in plasma jet properties when impinging on grounded or floating targets have been reported ~\cite{bhatt_cell_2014,gazeli_experimental_2020}. To generate APPJs most of them use sinusoidal voltages, ranging in frequencies from kHz to MHz.
Previous works of our groups revealed that using conducting or dielectric targets (or sample holders) has important effect not only on plasma jet properties but also on plasma treatment outcome ~\cite{prysiazhnyi_properties_2017, do_nascimento_changes_2017}. For instance it was verified that better adhesion of polydimethylsiloxane (PDMS) samples were achieved after plasma treatment using a conductive sample holder ~\cite{do_nascimento_changes_2017}.

In this work we studied the behavior of three main parameters of an APPJ: rotational and vibrational temperatures ($T_r$ and $T_v$, respectively) and mean discharge power ($P_{plasma}$), in a DBD device in two distinct configurations. The first one employed is more often reported in the literature, in which a plasma jet is produced between two dielectric barriers, one enveloping the powered electrode and the other covering the grounded one. In the second configuration the dielectric on the powered electrode was partly removed while the one covering the grounded electrode was kept. This last configuration was not extensively explored yet, mainly due to safety issues appearing in medical applications of APPJs. Some works using microwave, sinusoidal and pulsed power sources have been reportedly using it but without reporting measurements of plasma power, vibrational or rotational temperatures or comparison for different working gases ~\cite{lu_single_2008,wei_optical_2011,gessel_temperature_2013,gaens_reaction_2014,ilik_optical_2016, darny_analysis_2017,deepak_low_2017,qiao_temporal_2019,zhang_discharge_2019}. Since some applications may not require the safe operating conditions provided by the dielectric barrier over the powered electrode, these can take advantage of the higher plasma power and other jet properties that can be achieved when the dielectric barrier is removed.
 
\section{Materials and Methods}
\begin{SCfigure}[1][htb]
\centering
\includegraphics[width=12 cm]{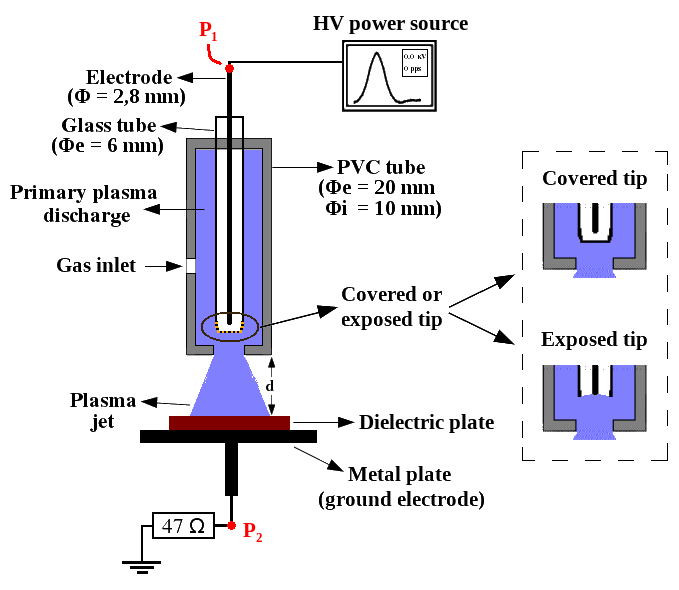}
\caption{Scheme used to produce the plasma jets in both configurations.\label{ExpSetup}}
\end{SCfigure}

A schematic layout of the experimental setup used in this work is shown in Fig. \ref{ExpSetup}. In order to produce the plasma jets using the pulsed power source, high voltage (HV) pulses of $\sim$20 kV amplitude, $\sim$400 ns width and 60 Hz repetition rate were applied to a pin electrode covered with a glass tube with one of its ends closed (covered tip) or open (exposed tip), for the double and the single barrier configurations, respectively. Details about the pulsed power source can be found in ~\cite{machida_ferrite_2015}. A biaxially-oriented polyethylene terephthalate (BoPET) foil, 500 µm-thick and 10 cm-sided, was placed on the top of a grounded electrode and acted as a common dielectric barrier in both cases. The distance $\textbf{d}$ between the end of the dielectric enclosure (a polyvinyl chloride tube) and the dielectric plate was kept constant and equal to 10 mm. Argon (Ar) and helium (He), both with 99.99$\%$ purity, were used as working gases at the same flow rate of 3.0 l/min.
The pulse-like power source consists of a HV supply whose output presents a damped sine waveform, with an oscillation frequency of $\sim$150 kHz and 60 Hz repetition rate. The maximum positive voltage peak provided by this power source is $\sim$15 kV. The other parameters used in the experiment with the pulse-like power source are the same as those mentioned for the pulsed one.

In order to obtain the rotational and vibrational temperature of $N_2$ molecules, we use spectroscopic emissions from the $N_2$ second positive system, $C \: {}^{3} \Pi_u, \nu' \rightarrow B \: {}^{3} \Pi_g , \nu''$ (referred as $N_2 (C \rightarrow B)$ hereafter), with $\Delta \nu$ = $\nu' - \nu''$ = -2) in the wavelength range from 360 to 385 nm ~\cite{moon_comparative_2003,bruggeman_gas_2014,zhang_determination_2015,ono_optical_2016}. Then comparisons between measured and simulated spectra are performed and temperature values are determined by those that generate simulated curves that best fit to the experimental spectra. The spectroscopic measurements were performed using a portable multi-channel spectrometer from OceanOptics (model HR4000), with FWHM equal to (0.545 $\pm$ 0.007) nm. The spectra simulations were performed using data from the SpecAir software ~\cite{specair_website_2020}. We defined the uncertainties in the temperature measurements as:

\begin{equation}
{\sigma T = \sqrt{(\Delta T/2)^2 + [(1- R^2 ) T]^2}}\label{eqerrtemp}
\end{equation}

\noindent where $R^2$ is the coefficient of determination obtained in the comparison between experimental and simulated spectra, $\Delta T$ is the temperature step used in the simulations, and $T$ is the temperature value obtained for $T_r$ or $T_v$.

The mean discharge power ($P_{plasma}$) calculations were made by measuring simultaneously the voltage applied on the powered electrode (point $P_1$ in the Fig. \ref{ExpSetup}) and the voltage drop across a serial resistor R = 47 $\Omega$ (point $P_2$ of Fig. \ref{ExpSetup}), which is used to calculate the current that flow through the plasma. In order to measure the applied voltage at $P_1$ a 1000:1 voltage probe (Tektronix model P6015A) was used, and the voltage measurement at $P_2$ was performed using a 100:1 voltage probe. The signals waveforms were recorded using a 100 MHz oscilloscope from Tektronix (model TBS1104B).
Then, the $P_{plasma}$ value is obtained through the integration of the product between voltage ($V(t)$) and current ($i(t)$) signals during the time of plasma pulse duration multiplied by the pulse repetition rate ($f$), that is ~\cite{holub_measurement_2012,ashpis_progress_2017}:

\begin{equation}
{P_{plasma} = f \int_{t_1}^{t_2} V(t) i(t) dt}\label{eqpower}
\end{equation}

Another important parameter that correlates optical measurements with an electrical quantity in APPJs is the reduced electric field strength $E_n = E/n$, where $E$ is the electric field strength and $n$ is the gas number density. The $E_n$ value can be estimated using the ratio between intensity emissions from $N_2^{+}$ ions and excited $N_2$ ($I_{N_2^{+}}/I_{N_2}$), and changes in this ratio indicate change in $E_n$, being that the higher the ratio $I_{N_2^{+}}/I_{N_2}$, the higher the $E_n$ value, that is $E_n \propto I_{N_2^{+}}/I_{N_2} $ ~\cite{paris_intensity_2005,zhang_discharge_2019}. Usually the emission from the first negative system of $N_2^{+}$ from the band ($B \: {}^{2} \Sigma_g{+}, \nu' = 0 \rightarrow X \: {}^{2} \Sigma_g^{+} , \nu''$ = 0), emitting at $\lambda $ = 391.4 nm together with a $N_2 (C \rightarrow B)$ emission coming from $N_2 (C)$ energy level with $\nu'$ = 0 or 2 are used to calculate the $I_{N_2^{+}}/I_{N_2}$ ratio and obtain $E_n$. We choose to use the $N_2 (C, \nu' = 0 \rightarrow B, \nu'' = 2)$, emitting at $\lambda$ = 380.49 nm to obtain $I_{N_2}$ (referred as $I_{380}$ hereafter) as well as the usual $N_2^{+}$ emission at $\lambda $ = 391.4 nm to obtain $I_{N_2^{+}}$ (referred as $I_{391}$ hereafter). However, the $I_{391}/I_{380}$ ratios were not used to obtain $E_n$ values in the APPJs in this work, but were used to evaluate possible changes in plasma jet behaviors when switching from the double to the single barrier configuration.

In order to perform the tests attempting to remove copper (Cu) films deposited onto a 99.9$\%$ purity polished alumina (Al$_2$O$_3$) substrate, one square sample 1 inch-sided was used. The thickness of Cu film is approximately 500 nm. The plasma application was performed statically, that is, the sample was positioned under the plasma jet and was not moved during the entire application time interval, which was equal to five minutes in each case. This part of the work was performed using only the pulsed power source to produce the plasma jets, since preliminary tests did not revealed significant removal of Cu film when using the pulse-like source.

\section{Results}

\subsection{Electrical measurements}
\begin{SCfigure}[1][htb]
\centering
\includegraphics[width=10 cm]{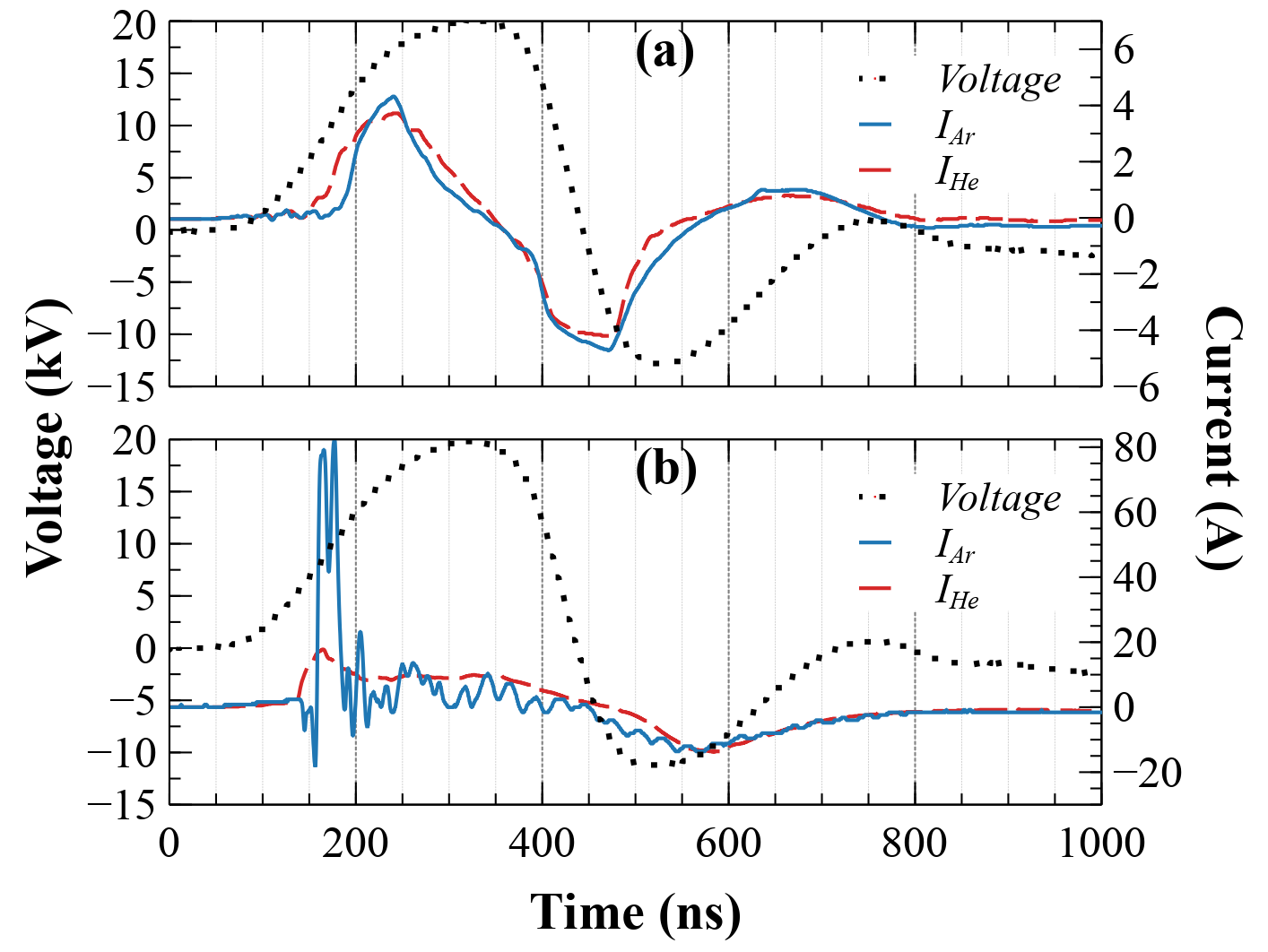}
\caption{Current and voltage waveforms measured in (a) double barrier configuration and (b) single barrier using the pulsed voltage source.\label{pulsedwf}}
\end{SCfigure}

Figure \ref{pulsedwf} shows the measured current waveforms obtained using the pulsed source with Ar and He as working gases for (a) the double barrier configuration and (b) the single barrier one. Typical high-voltage (HV) waveforms obtained in each case, which have good repeatability, are also shown in Fig. \ref{pulsedwf}. The values of the current measured in the single barrier configuration are notably higher, which is in agreement with what is expected to happen without an insulating barrier, justifying the choice of a double dielectric barrier configuration for applications that require electrical safety. The values obtained for the power in the double barrier configuration were 0.62 W when using Ar as the working gas, and 0.64 W when using He. For the single barrier configuration the power values were 2.98 W using Ar and 3.29 W using He. By changing the device configuration from double barrier to single barrier the $P_{plasma}$ values increased dramatically by approximately five times for both working gases, which is a great advantage for applications that require higher discharge power.

An interesting observation about the current waveforms shown in Fig. \ref{pulsedwf} is that when operating in the double barrier configuration, the current signals for Ar and He gas (solid blue and dashed red curves, respectively) are almost equal, while in the single barrier configuration the current curves obtained for different gases do not behave in the same way, which suggests that different regimes are taking place depending on the working gas for the latter case. It can be also noticed that in the single barrier case, despite the $P_{plasma}$ values being close, the current signal measured with Ar ($I_{Ar}$) presents a very high peak value ($\sim$80 A) at the beginning of the discharge, that is nearly four times higher than the peak value of the current obtained with He ($I_{He}$). Also, the temporal behavior of $I_{Ar}$ is not as smooth as that observed for $I_{He}$, being $I_{Ar}$ that presents a lot of oscillations as time evolves, which is another indication that the regime of the plasma jet using Ar as the working gas changes when the barrier over the powered electrode is removed. However, the same does not seem to happen when He is used as the working gas, which exhibits quite similar current behavior in both configurations.

In Fig. \ref{pulsedwf} we can also see that in the single barrier configuration the discharge currents start increasing earlier in the time interval between 100 and 150 ns, while using the double barrier the current peaks begin growing between 150 and 200 ns. In other words, as expected lower voltage values are required to ignite the discharge in the single barrier case due to the powered electrode being in contact with the working gas and thus releasing more electrons from the metal to the plasma. Those findings can also be the reasons for the more extended duration of the plasma discharge, depicted by the wider current pulses observed comparing figures \ref{pulsedwf} (a) and (b).

\begin{SCfigure}[1][hbt]
\centering
\includegraphics[width=10 cm]{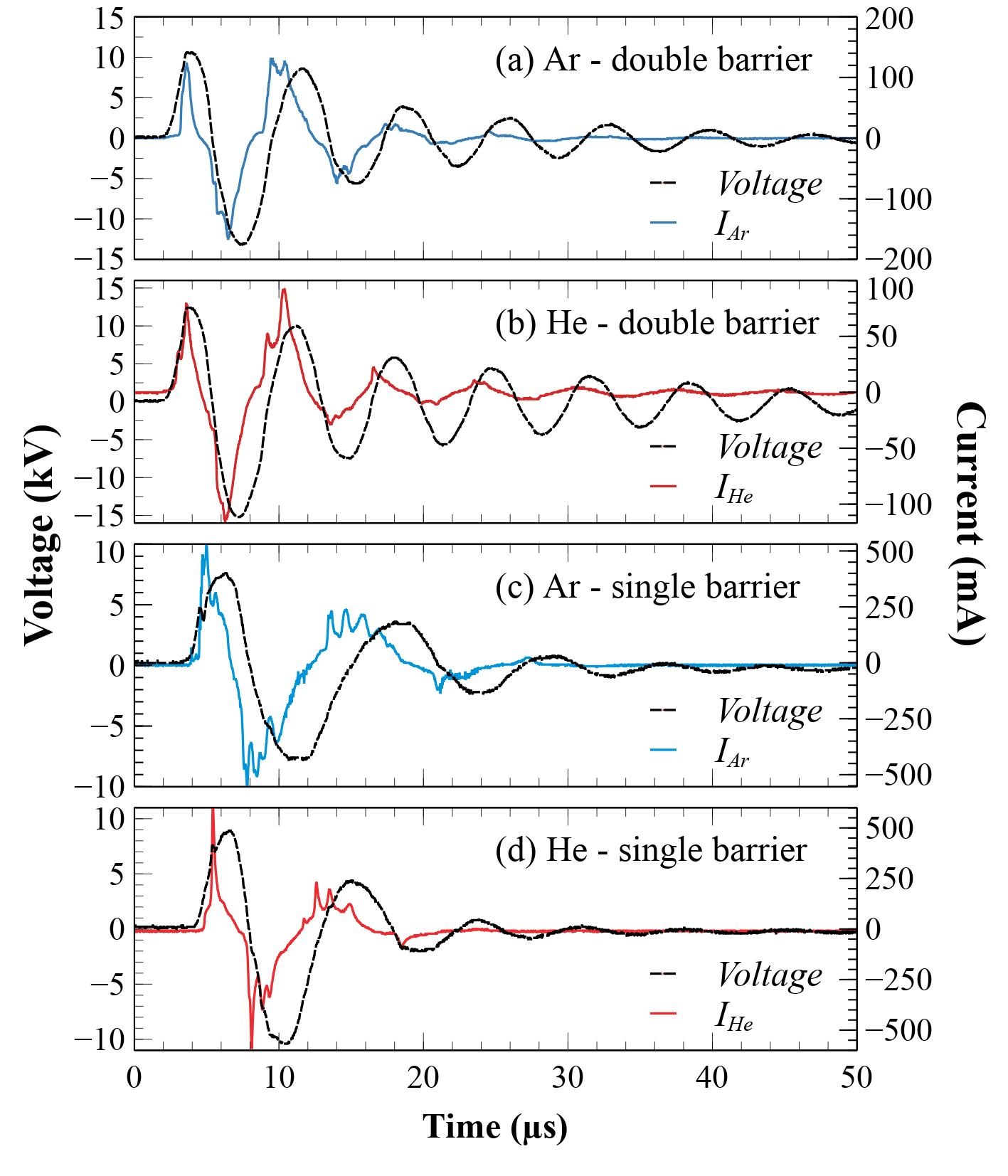}
\caption{Voltage and current waveforms measured using double barrier for Ar (a) and He (b), and using single barrier for Ar (c) and He (d) using the pulse-like voltage source.\label{plikedwf}}
\end{SCfigure}

Figure \ref{plikedwf} shows the measured voltage and current waveforms obtained in double barrier configuration for Ar (a) and He (b), and in single barrier mode for Ar (c) and He (d) using the pulse-like voltage source. A point that is noteworthy when comparing double and single barrier configurations using the same gas (Ar or He) is that, unlike the pulsed power source, when the pulse-like source is used significant voltage drops, from 10-12 kV to $\sim$8-9 kV for the first positive peaks, are observed when changing the configuration. The occurrence of that voltage drop is due to the fact that the pulse-like source is more sensitive to the impedance matching, which can be verified comparing the frequency of the damped sine voltage, which are $\sim$150 kHz and $\sim$110-120 kHz in the double and single barrier configurations, respectively.

The values obtained for $P_{plasma}$ calculated using data presented in Fig. \ref{plikedwf} were 0.27 W when using Ar as the working gas, and 0.23 W when using He in the double barrier configuration and 0.49 W using Ar and 0.48 W using He in the single barrier mode. Despite the voltage drop observed when changing the double barrier configuration to the single one, the observed values for the power increased in the second case. However, the increase was only $\sim$1.8 times using Ar and $\sim$2.1 times using He.

\subsection{Spectroscopic measurements}

\begin{figure}[htb]
\centering
\begin{minipage}[t]{0.48\textwidth}
\includegraphics[width=\textwidth]{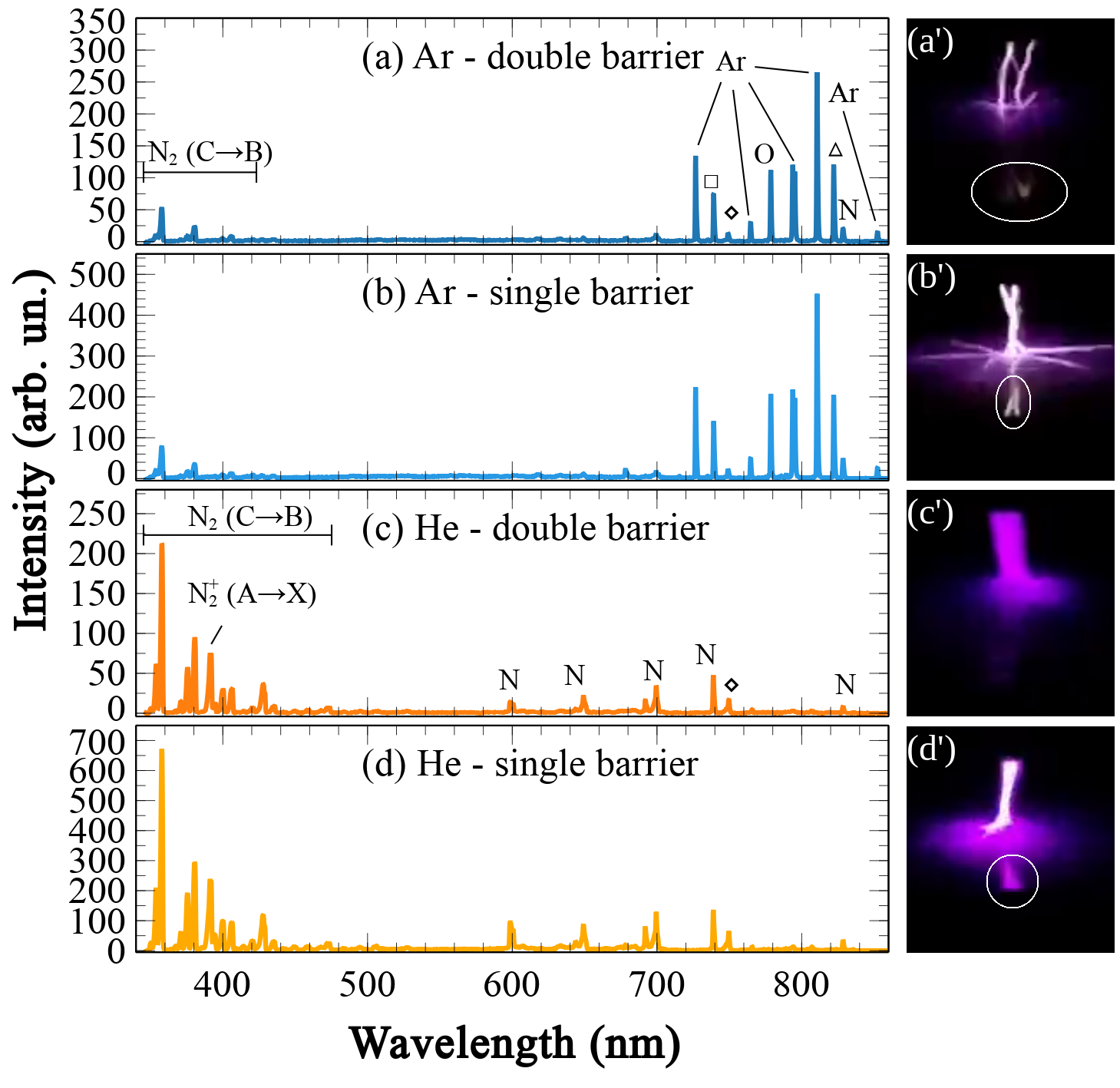}
\caption{Emission spectra obtained using the pulsed power source with argon as the working gas in (a) and (b) for the double and single barrier configurations, respectively, and the same for using helium in (c) and (d). (a')-(d') are the corresponding photos of plasma jets. The ellipses indicate reflections.}\label{pulsespec}
\end{minipage}
\hfill
\begin{minipage}[t]{0.48\textwidth}
\includegraphics[width=\textwidth]{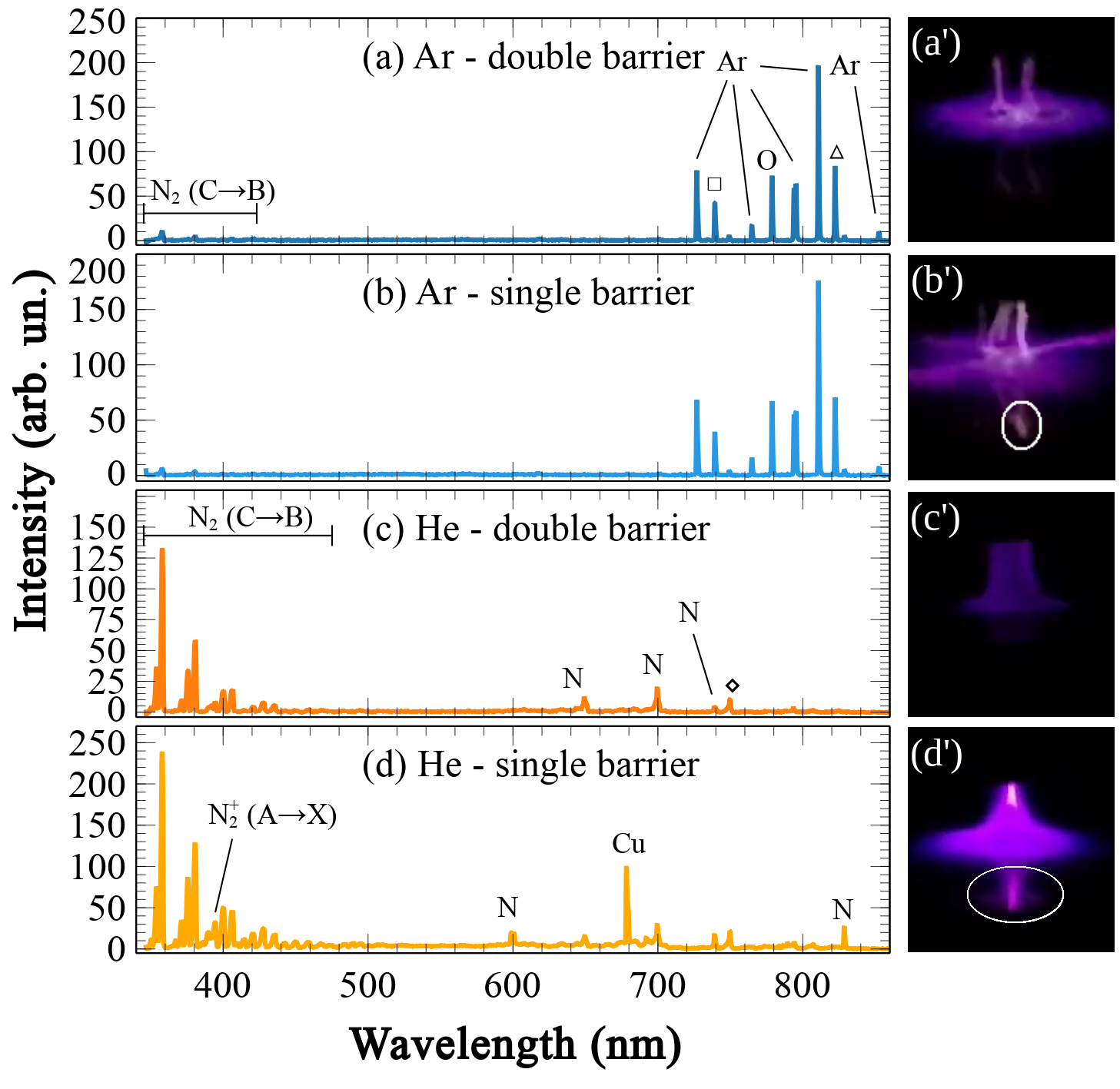}
\caption{Emission spectra obtained using the pulse-like power source with argon as the working gas in (a) and (b) for the double and single barrier configurations, respectively, and the same for using helium in (c) and (d). (a')-(d') are the corresponding photos of plasma jets. The ellipses indicate reflections.}\label{plikespec}
\end{minipage}
\end{figure}

\begin{figure}[htb]
\centering
\begin{minipage}[b]{0.49\textwidth}
\includegraphics[width=\textwidth]{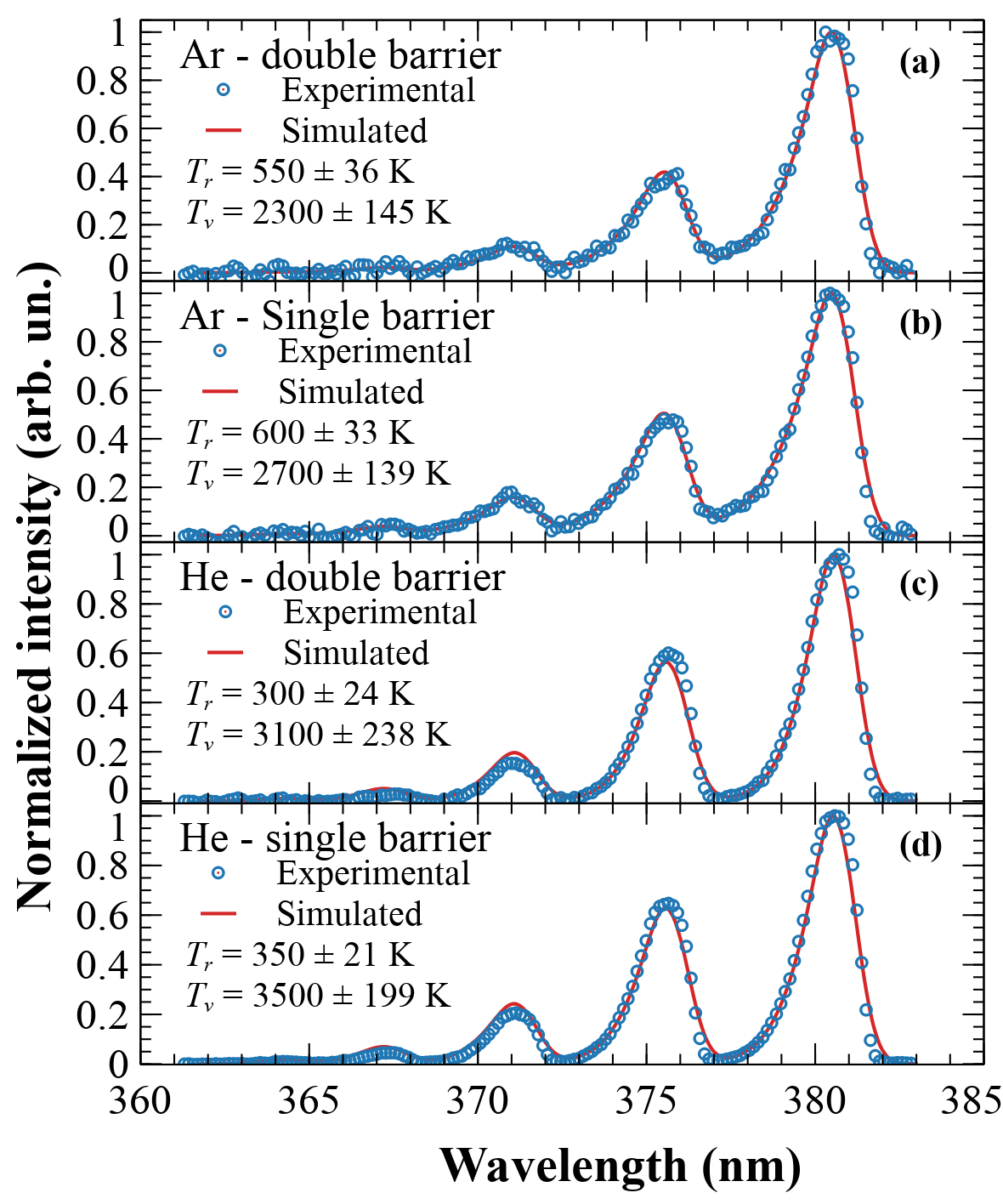}
\caption{Details of measured $N_2 (C \rightarrow B)$ band emissions (circles) in the wavelength range from 360 to 385 nm and the respective simulated spectra (red lines) that best fit the experimental data using the pulsed power source.}\label{pulsedn2}
\end{minipage}
\hfill
\begin{minipage}[b]{0.49\textwidth}
\includegraphics[width=\textwidth]{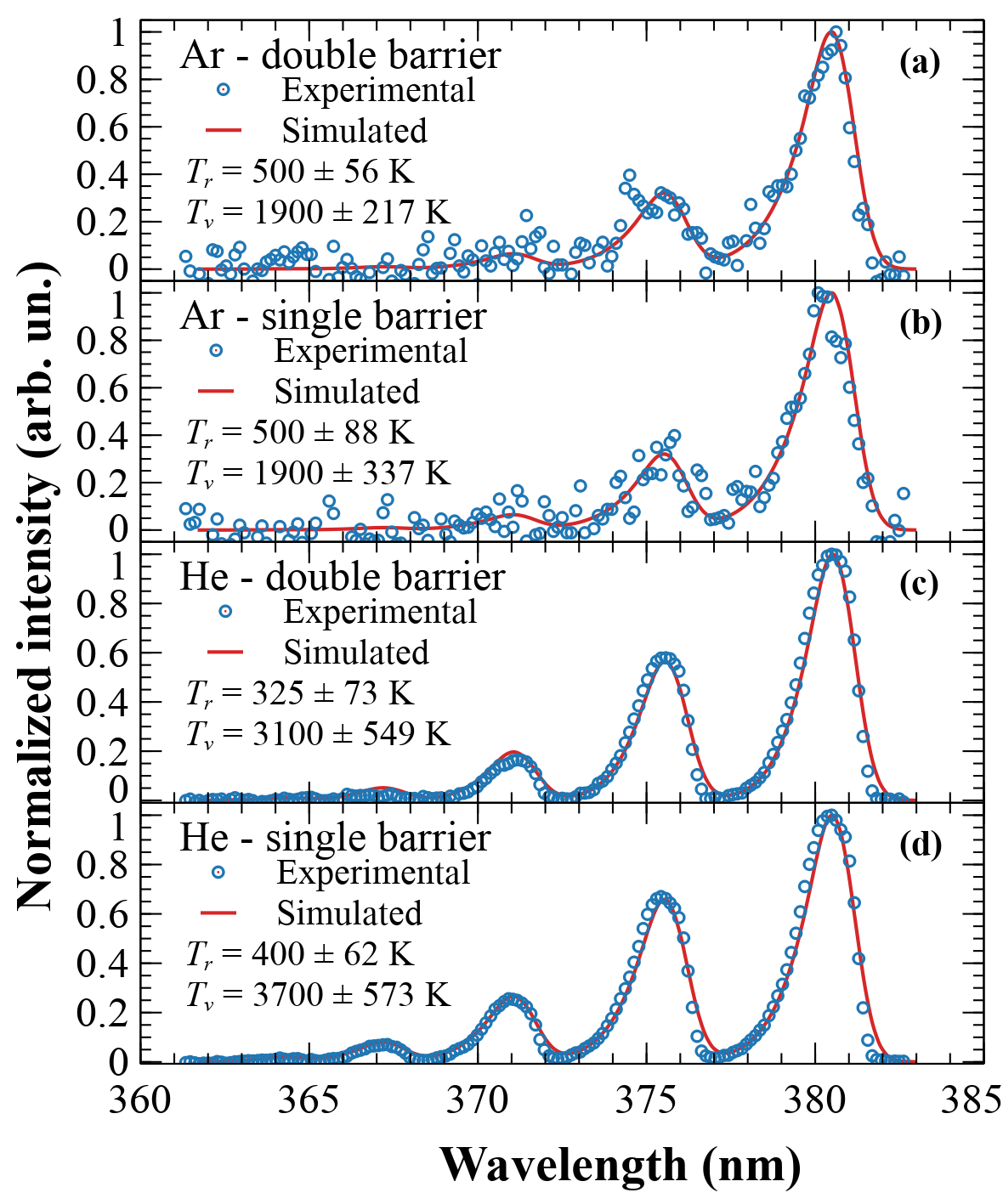}
\caption{Details of measured $N_2 (C \rightarrow B)$ band emissions (circles) in the wavelength range from 360 to 385 nm and the respective simulated spectra (red lines) that best fit the experimental data using the pulse-like power source.}\label{pliken2}
\end{minipage}
\end{figure}

An overview of the emission spectra in all conditions studied in this work is shown in Fig. \ref{pulsespec}, for the pulsed source, and in Fig. \ref{plikespec} for the pulse-like one. All atomic emissions shown in Figs. \ref{pulsespec} and \ref{plikespec} are from neutral species in excited states. Squares indicate superposed emissions from Ar and N, triangles indicate superposition of O and N lines and diamonds indicate an emission from the first positive system of $N_2$. On the right side of Figs. \ref{pulsespec} and \ref{plikespec} are also shown photos of the corresponding plasma jets produced in each configuration/working gas. The detailed views of the $N_2 (C \rightarrow B)$ emission band used to calculate $T_r$ and $T_v$ are shown in Figs. \ref{pulsedn2} and \ref{pliken2}, for the pulsed and pulse-like power sources, respectively.

Comparing the spectra in the Fig. \ref{pulsespec} obtained using the pulsed source with double or single barriers for each gas, the noticeable differences are that the intensities of the atomic and molecular emissions are more intense when the plasma jet is operated using single dielectric barrier, which is in an agreement with the higher apparent luminosities shown in the corresponding photos. Furthermore, the change from double to single barrier plasma jet configuration do not result in any new excited species nor species emitting radiation in other wavelengths, that is, no new excitation levels were observed after the change. It is interesting to notice that the relative intensities of atomic species remain almost the same in the different configurations. However, even though there are no significant changes in the relative emissions intensities of different species, the fact that they have increased in absolute values when changing from the double barrier to the single barrier configuration indicates the production of more exited reactive species. This observation is associated to the higher discharge power values observed when using the single barrier configuration, which is also in agreement with studies reported on the literature ~\cite{baek_effects_2016,do_nascimento_four-electrodes_2020}.
On the other hand, it was verified that the molecular species alone presented significant modifications in their relative intensities when keeping the same emission band (same $\Delta \nu $), and this results in significant changes on the $T_v$ values, as can be seen in Fig. \ref{pulsedn2}. 

Contrary to what happened with the pulsed source when switching from the double to the single barrier configuration, when using the pulse-like power source it was observed an increase in the intensity of the emissions from atoms and molecules only when He was used as the working gas. An interesting observation can be made on the appearance of an emission line whose peak was detected at 678.29 nm in Fig. \ref{plikespec}(d), which is possibly an emission from neutral Cu atoms ($\lambda_{Ritz}$ = 677.57 nm). However, since there are no other line emissions from Cu in the observed spectrum, further investigation will be required in order to confirm this information.

From the spectra shown in Fig. \ref{pulsespec}, obtained from the plasmas produced using the pulsed power source, one can see that it is possible to calculate the $I_{391}/I_{380}$ ratios only when He is used as the working gas because $N_2^{+}$ emissions are not present in the spectra obtained using Ar to generate the plasma jets. From Figs. \ref{pulsespec}(c) and \ref{pulsespec}(d), the $I_{391}/I_{380}$ ratios for the double and single barrier configurations are 0.786 and 0.811, respectively, which corresponds to an increase of $\sim 3 \% $ in the $E_n$ value when switching from the double to the single barrier configuration. In other words, $E_n$ remains almost constant when switching the device configuration, which means that the observed increase in the number of emitting species shown in Fig. \ref{pulsespec}, caused by the increase in the electric field strength ($E$) that occurs when the dielectric barrier between the powered electrode and the plasma is removed, is related only to the intensity of $E$.
Since $E_n$ is almost constant when we switch from the double barrier configuration to the single one, it suggests that there are no changes in the operating plasma regime associated with the change in the device configuration when He is used as the working gas. It was not possible to perform the same analysis using the pulse-like power source due to the low intensity emission from the $N_2^{+}$ measured using the double barrier configuration.

From the photos shown in Figs. \ref{pulsespec}(a'-d') and \ref{plikespec}(a')-(d'), one can notice that the plasma jets produced using the single barrier configuration tend to spread further on the impinged surface than those produced in the double barrier configuration (compare a' with b' and c' with d'). These differences in the size of plasma jets spreading on the surface are related to the different electrical potentials (higher without the barrier on the powered electrode and smaller when the second dielectric barrier is used) that are being applied to the plasma in each condition. Therefore in the single barrier APPJ configuration, the potential difference between grounded electrode and the plasma is higher, producing higher current as well with a greater probability for the plasma plume to reach the target.

From Fig. \ref{pulsedn2} we can see that when the pulsed power source is used both $T_r$ and $T_v$ values change when switching from the double barrier configuration to the single one, for both working gases used. Usually in APPJs the $T_r$ value is considered to be very close to the gas temperature ($T_{gas}$), that is $T_r$ $\approx$ $T_{gas}$. Therefore regarding the obtained small variations in $T_r$ values, 50 K for both gases, it is a good finding, since plasma jets as cold as possible are desirable in order to avoid possible thermal damages on samples subject to APPJ treatment. In relation to the increment in $T_v$ values for the single barrier plasma jet, 400 K for both gases, it is a very good result because one wants to produce APPJs with $T_v$ values as high as possible due to the relationship between this parameter and chemical reactions rate. Thus plasma jets with higher $T_v$ values are able to induce higher degree of surface activation ~\cite{ lambert_vibrationvibration_1967,smith_preference_2004,do_nascimento_comparison_2017}. On the other hand, as can be seen in Fig \ref{pliken2} the temperature variations obtained using the pulse-like source can not be considered significant since the uncertainties in the temperature values are higher than 15$\%$ of the measured ones in all cases. However, the increase in $T_r$ and $T_v$ values obtained using He when switching from the double to the single barrier configuration can be considered as a trend.


\subsection{Applications on removal of copper films deposited onto alumina substrates}
\begin{figure}[htb]
\centering
\includegraphics[width=15 cm]{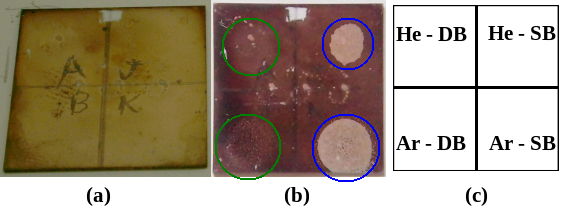}
\caption{Photos of Cu films deposited onto Al$_2$O$_3$ substrates before (a) and after (b) plasma application using the pulsed power source. (c) Regions where plasma jets were applied using He and Ar in both configurations. \label{cufilms}}
\end{figure}

In order to verify some effects related to the choice of using the powered electrode covered or not with a dielectric material, we applied the plasma jets on a copper (Cu) film deposited onto an alumina (Al$_2$O$_3$) surface using the pulsed power source.
A consideration that should be taken into account before analyse the results is that the sample is a conducting material. So, it acts as a floating electrode when the plasma jet is applied to it and when using the powered electrode without the dielectric barrier, the discharge plasma regime can not be considered a DBD discharge. However, since the grounded electrode still covered with a dielectric, it is a valid experiment concerning the use of a powered electrode covered or not with a dielectric material.
The sample used to perform the tests is shown in Fig. \ref{cufilms}(a). Figure \ref{cufilms}(b) shows the visual effects after apply on the Cu film plasma jets formed using each of the configurations studied in the previous sections, for five minutes each case. 
The regions where the plasma jets were applied using Ar or He as working gases in double barrier (DB) or single barrier (SB) configurations are indicated in Fig. \ref{cufilms}(c).

Comparing the photos in Figs. \ref{cufilms}(a) and \ref{cufilms}(b) it can be seen that evident visual effects are observed only when the single barrier configuration was used, operating with both Ar and He gases. The rounded marks in the sample, highlighted in blue in Fig. \ref{cufilms}(b), correspond to the regions where there was significant removal of Cu film while in the regions highlighted in green, the removal of film was less significant. In preliminary tests we had found that the use of Ar in the single barrier configuration promoted greater removal of Cu films. Thus, we chose to apply plasma to the sample in the following order Ar-SB, He-SB, Ar-DB and He-DB. So, the possible effects of plasmas applied in one region to affect the other would favor greater removal of Cu film with less potent plasma conditions. However, this precaution is a redundancy, because when impinging the surface of the Cu film, the plasma does not spread through it as it would happen with an insulating target. The observed change in copper color before and after applying plasma is only due to the differences in lighting used to take the photos.

From Fig. \ref{cufilms}(b) we can also verify that in the single barrier configuration the use of Ar as the working gas was able to remove the Cu film over an area larger than when using He. This result may be related to the higher peak current that occurs when using Ar.

\section{Discussion and conclusions}

The experimental results obtained in this work using the pulsed power source are summarized in Table \ref{table1}, while the results obtained using the pulse-like source are summarized in Table \ref{table2}. 

Table \ref{table1} shows that the values of all measured parameters increase when changing from the double barrier configuration to the single one, using Ar or He as working gases. As can be seen in Table \ref{table1}, the $T_r$ and $T_v$ values presented differences not so far from the measurements uncertainties (especially for $T_r$). On the other hand, changing from double to single barrier configuration the $P_{plasma}$ values increased about 4.8 and 5.1 times when using Ar and He gases, respectively. This last result was expected because when the powered electrode is in contact with the working gas, a higher electric current will flow through the plasma jet, a result that is confirmed by the current waveforms measured for Ar and He using the double or the single barrier configurations. 

\begin{table}[htb]
\caption{Summarized results for the measured plasma temperatures and plasma power using the pulse voltage.\label{table1}}
\centering
\begin{tabular}{ccccccc}
\hline
\textbf{Parameter} & \multicolumn{2}{c}{\textbf{$T_{r}$ (K)}} & \multicolumn{2}{c}{\textbf{$T_{v}$ (K)}} & \multicolumn{2}{c}{\textbf{$P_{plasma}$ (W)}} \\ 
\textit{\textbf{Working gas}} & \textit{Ar} & \textit{He} & \textit{Ar} & \textit{He} & \textit{Ar} & \textit{He} \\ 
\hline
Double barrier & 550 $\pm$ 36 & 300 $\pm$ 24 & 2300 $\pm$ 145 & 3100 $\pm$ 238 & 0.62 & 0.64 \\ 
Single barrier & 600 $\pm$ 33 & 350 $\pm$ 21 & 2700 $\pm$ 139 & 3500 $\pm$ 199 & 2.98 & 3.29 \\ 
\hline
\end{tabular} 
\end{table}

Concerning the different $T_r$ and $T_v$ values obtained in the two DBD configurations, one can conclude that it would be beneficial to use the single barrier one, since $T_r$ does not change significantly while the $T_v$ is higher compared to the double barrier case. However, due to the higher current and power values obtained for the single barrier jet, it is not readily suitable for applications that require electrical safety for the device operator or the target impinged by the plasma jet, as in \textit{in vivo} applications. Nevertheless, the single barrier configuration is quite attractive for treatments of materials that require more powerful plasma to achieve an adequate degree of surface activation or higher interaction between the plasma and the target. An example of that is the application reported by Gazeli \textit{et al} ~\cite{gazeli_experimental_2020}, where it was verified that a more powerful plasma jet is more efficient in the removal of resistant bibenzyl deposits formed on glass surfaces.

Once the $T_r$ and $T_v$ values do not increase in the same proportion as $P_{plasma}$ increases when changing the DBD configuration, one can infer that there is no linear dependency between discharge power and the plasma jet temperatures, which is in agreement with some results reported in the literature ~\cite{moon_uniform_2004}. However, to confirm or not this trend a further investigation is necessary. Besides, the possibility of plasma jets operating in different regime depending on the working gas for the single barrier case also needs to be investigated in more detail. In addition, studies using sinusoidal voltages instead of pulsed ones to produce the plasma jets should be performed in future experiments to complement the understanding of the differences in the plasma jet properties that occur when the device configurations are changed.

In Table \ref{table2} we can see that when the pulse-like power source was used only the $P_{plasma}$ parameter changed significantly when switching the configuration from the double barrier to the single one. However, the observed increase in $P_{plasma}$ did not occur in the same proportion observed when using the pulsed source. This can be attributed partially to the voltage drop that occurred due to the change in electrode configuration and partially to the reduction of the observed frequency in the damped-sine waveform. Nevertheless, we can speculate that a pulsed voltage presents a more effective way for energy transfer from the power source to the plasma when operating in the single barrier mode.

\begin{table}[tb]
\caption{Summarized results for the measured plasma temperatures and plasma power using the pulse-like voltage.\label{table2}}
\centering
\begin{tabular}{ccccccc}
\hline
\textbf{Parameter} & \multicolumn{2}{c}{\textbf{$T_{r}$ (K)}} & \multicolumn{2}{c}{\textbf{$T_{v}$ (K)}} & \multicolumn{2}{c}{\textbf{$P_{plasma}$ (W)}} \\ 
\textit{\textbf{Working gas}} & \textit{Ar} & \textit{He} & \textit{Ar} & \textit{He} & \textit{Ar} & \textit{He} \\ 
\hline
Double barrier & 500 $\pm$ 56 & 325 $\pm$ 58 & 1900 $\pm$ 217 & 3100 $\pm$ 549 & 0.27 & 0.23 \\ 
Single barrier & 500 $\pm$ 88 & 400 $\pm$ 62 & 1900 $\pm$ 337 & 3700 $\pm$ 573 & 0.49 & 0.48 \\ 
\hline
\end{tabular} 
\end{table}


\section*{Acknowledgements}F.N. thanks the Center for Semiconductor Components and Nanotechnologies (CCSNano/UNICAMP) for the loan of some equipments used in this work. This research received financial support from Brazilian National Council for Scientific and Technological Development (CNPQ) and from São Paulo Research Foundation (FAPESP) under grant $\#$2020/09481-5.

\bibliographystyle{unsrt}
\bibliography{references-bibtex}

\end{document}